# Multi-kV class β-$Ga_2O_3$ MESFETs with a Lateral Figure of Merit up to 355 MW/cm$^2$

Arkka Bhattacharyya, Praneeth Ranga, Saurav Roy, Carl Peterson, Fikadu Alema, George Seryogin, Andrei Osinsky and Sriram Krishnamoorthy

***Abstract*—We demonstrate over 3 kV gate-pad-connected field plated (GPFP) β-$Ga_2O_3$ lateral MESFETs with high lateral figure of merit (LFOM) using metalorganic vapor phase epitaxy (MOVPE) grown channel layers and regrown ohmic contact layers. Using an improved low-temperature MOVPE selective area epitaxy process, we show that a total contact resistance to the channel as low as 1.4 Ω.mm can be achieved. The GPFP design adopted here using plasma-enhanced chemical vapor deposited (PECVD) $SiN_x$ dielectric and $SiN_X/SiO_2$ wrap-around passivation exhibits up to ~14% improved $R_{ON}$, up to ~70% improved breakdown voltage ($V_{BR}$ = $V_{DS}$-$V_{GS}$) resulting in up to 3× higher LFOM compared to non-FP β-$Ga_2O_3$ lateral MESFETs. The $V_{BR}$ (~2.5 kV) and LFOM (355 MW/cm$^2$) measured simultaneously in our GPFP β-$Ga_2O_3$ lateral MESFET (with $L_{GD}$ = 10μm) is the highest value achieved in any depletion-mode β-$Ga_2O_3$ lateral device to date.**

## I. Introduction

$Ga_2O_3$-based devices have increasingly gained momentum as a potential technology that demonstrates numerous strengths for high voltage/power applications. The high projected breakdown field (~8 MV/cm) of β-$Ga_2O_3$ and the availability of high-quality melt-grown β-$Ga_2O_3$ bulk substrates offers compelling promise for high efficiency power devices[1]–[6]. The β-$Ga_2O_3$–based device performance has rapidly progressed in both lateral and vertical geometries thanks to the simultaneous progress in high-quality epilayer growth technologies that includes mainly molecular beam epitaxy (MBE), metalorganic vapor phase epitaxy (MOVPE), halide vapor phase epitaxy (HVPE) and low-pressure chemical vapor deposition (LPCVD) as well as device processing techniques [7]–[16]. Of all these epitaxial growth techniques, MOVPE has captivated widespread attention due to its versatility and ability to grow comparatively higher-quality epilayers that enables high room-temperature electron mobility values (close to the theoretical limit) and could be promising for fabricating $Ga_2O_3$ lateral FETs with high current densities as well as high breakdown voltages [8], [17]–[19].

Many new field management techniques have been successfully demonstrated in lateral β-$Ga_2O_3$-based devices that enabled high-breakdown voltages, high average breakdown fields, and high ($V_{BR}^2/R_{on,sp}$) lateral figures of merit but not simultaneously in the same device [3], [4], [20]. In other words, $V_{BR}$ >2kV, lateral $V_{BR}^2/R_{on,sp}$ > 300MW/cm$^2$, and $E_{BR,AVG}$ > 2MV/cm could not be achieved simultaneously in the same device. This could be due to a combination of factors such as device design, device processing techniques (bulk or surface leakage paths) and more importantly the epilayer material quality. In this work, we address these issues using a gate-pad connected field plate (GPFP) design using a PECVD deposited $SiN_x$ dielectric in a β-$Ga_2O_3$ MESFET with MOVPE-grown channel and contact layers.

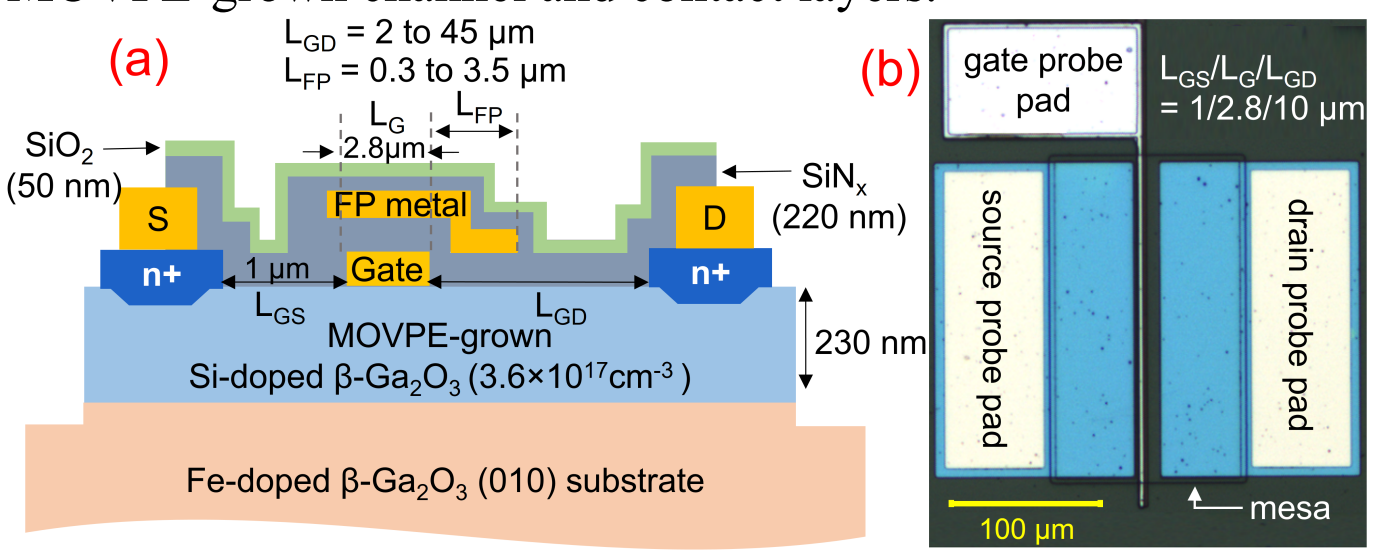

Fig. 1. (a) Schematic of the GPFP β-$Ga_2O_3$ MESFET with a $SiN_X/SiO_2$ wrap-around passivation. The FP metal is electrically connected to the gate pad outside the mesa (not shown). (b) top-view optical image of a fabricated GPFP MESFET.

## II. Device Growth and Fabrication

The epitaxial structure shown in Fig. 1(a) consists of a β-$Ga_2O_3$ channel (230 nm thick Si-doped ~3.6×10$^{17}$ cm$^{-3}$) on a Fe-doped (010) bulk substrate grown using Agnitron Technology's Agilis 700 MOVPE reactor with TEGa, $O_2$, and silane ($SiH_4$) as precursors and argon as carrier gas. The 10×15 mm$^2$ edge-defined film fed grown (EFG) semi-insulating Fe-doped (010) $Ga_2O_3$ bulk substrate was acquired from Novel Crystal Technology (NCT), Japan. Before the channel layer growth, the substrate was cleaned using HF for 30 mins. From Hall measurement, the channel charge and mobility were measured to be 5.7×10$^{12}$ cm$^{-2}$ and 95 cm$^2$/Vs respectively, yielding a channel sheet resistance, $R_{sh,ch}$ = 11.7 kΩ/□. The device mesa isolation and the source/drain ohmic contacts were selectively regrown by MOVPE technique using a Ni/$SiO_2$ mask pattern [21]. The contact recess etch was performed using a low power $SF_6$/Ar ICP-RIE dry etching, followed by a quick dip in a diluted BOE solution. The etching

This material is based upon work supported by the II-VI foundation Block Gift Program 2020-2021 and the *DoD SBIR Phase I – AF203-CS01 (Contract #: FA864921P0304)*. This work was performed in part at the Utah Nanofab sponsored by the College of Engineering and the Office of the Vice President for Research (The University of Utah).

A. Bhattacharyya, P. Ranga, S. Roy, C. Peterson and S. Krishnamoorthy are with the Department of Electrical and Computer Engineering, University of Utah, Salt Lake City, Utah, USA 84112.(email: sriram.krishnamoorthy@utah.edu)

Sriram Krishnamoorthy is also with Materials Department, University of California, Santa Barbara, California, USA 93106.(email: sriramkrishnamoorthy@ucsb.edu)

F. Alema, G. Seryogin and A. Osinsky are with Agnitron Technology Incorporated, Chanhassen, Minnesota 55317, USA.

* Corresponding author e-mail: a.bhattacharyya@utah.edu

in the contact regions was extended down to the $Ga_2O_3$ epitaxial layer with an estimated $Ga_2O_3$ trench depths of 10-20 nm ($Ga_2O_3$ etch rate ~ 1.5 nm/min). The Si-doping in the regrown n$^+$ layer was ~ 2.6×10$^{20}$ cm$^{-3}$. Following the contact regrowth process, ohmic metal stack Ti/Au/Ni (20 nm/100 nm/30 nm) was evaporated on the regrown contact regions using photolithography patterning and lift off, followed by a 450°C anneal in $N_2$ for 1.5 mins. Ni/Au/Ni (30 nm/100 nm/30 nm) metal stack was then evaporated to form the Schottky gate for the MESFET structure.

The gate field plate design used in this work is shown in Fig.1(a), where the gate field plate metal was electrically connected to the gate pad (shorted) outside the device mesa, hence, named gate-pad-connected field plate (GPFP). This design was adopted to protect the channel region from the dry-etching plasma damage that occurs in the conventional gate field plate etch process flow and has been reported to cause plasma-damage induced $R_{ON}$ increase [4], [22]. Unlike MOSFETs, due to the absence of any gate dielectric that can act as etch stop/protective layer, avoiding the etch step in the active region played a key role to maintain $R_{ON}$ in our FP MESFETs as discussed later. The gate field plate metal was deposited following a $SiN_x$ (170 nm thick) passivation layer deposited using plasma-enhanced chemical vapor deposition (PECVD at 300°C). The field plate extension ($L_{FP}$) was varied from ~ 0.3 to 3.5 µm as the gate-to-drain distance ($L_{GD}$) varied from 2 to 45 µm. Finally, the whole active region was passivated using an $SiN_x/SiO_2$ (50 nm/50nm) bilayer dielectric deposited using the same PECVD (300°C) technique. The $SiN_x/SiO_2$ wrap-around bi-layer allowed for a thicker overall dielectric passivation with lower $SiN_x$ thickness to prevent excessive stress in the $SiN_x$ film. The final device optical image is shown in Fig.1(b). The non-FP MESFET devices had no dielectric surface passivation.

## III. RESULTS AND DISCUSSIONS

Fig. 2(a) & 2(b) show the DC output and transfer curves for the GPFP MESFET (solid lines) and the unpassivated non-FP MESFET (dashed lines) for device with dimensions $L_{GS}/L_G/L_{GD}$ = 1.0/2.8/2.4 µm. No hysteresis was observed in the DC current-voltage dual sweeps. The $I_{D,MAX}$ (at $V_{GS}$ = 0V) measured was 88 mA/mm for the passivated GPFP MESFET and 81 mA/mm for the un-passivated non-FP MESFET. The ON resistance ($R_{ON}$) @ $V_{GS}$ = 0V extracted from the linear region of the output curves were found to be 55.8 Ω.mm for the GPFP MESFET and 63.2 Ω.mm for the non-FP MESFET. From TLM measurements, the total $R_C$ (contact resistance) measured was 1.4 Ω.mm (< 3% of the total device $R_{ON}$) as shown in Fig. 2(c). This total $R_C$ value consists of metal/n$^+$ regrown $Ga_2O_3$ interface contact resistance, resistance of the regrown n$^+$-$Ga_2O_3$ region and the n$^+$ regrown $Ga_2O_3$/lightly doped $Ga_2O_3$ channel interface resistance. Compared to our previous work, the ten-fold improvement in $R_C$ was achieved using a low etch rate $SF_6$/Ar dry etching during the contact region recessing [21]. The transfer curves show that both devices have low leakage (~10$^{-12}$ A/mm) and high $I_{ON}/I_{OFF}$ ratio ~10$^{10}$. The passivated GPFP MESFETs show ~14% lower $R_{ON}$, ~8% higher ON currents, and ~13% higher transconductance compared to the non-FP MESFETs clearly indicating the FP design used here not only preserves the $R_{ON}$ in these devices, rather improves it possibly due to reduction of upward surface band-bending in the open access regions due to the $SiN_x$ passivation [23].

Fig. 2(d) shows the three-terminal breakdown characteristics (at $V_{GS}$ = -20 V) for the non-FP MESFETs and the GPFP MESFETs with various $L_{GD}$ values. All the breakdown measurements were performed with the wafer submerged in FC-40 Fluorinert dielectric liquid. Comparing devices with identical $L_{GD}$ values, the GPFP MESFETs show significant improvement in breakdown voltage ($V_{BR} = V_{DS} - V_{GS}$) over the non-FP MESFETs. The highest measurable $V_{BR}$ recorded was 2462V ($L_{GD}$ = 10µm) for the GPFP MESFET, which is ~70% higher compared to the non-FP MESFET (1462V) of the same dimension showing the efficacy of the FP design demonstrated here. The reverse leakage current rise was mainly due to gate leakage for both types of MESFETs. Fig. 2(e) and 2(f) show the OFF-state gate and drain terminal currents of the non-FP and GPFP MESFETs respectively with $L_{GD}$ of 10 µm, showing that the leakage current is dominated by the gate leakage before the catastrophic breakdown.

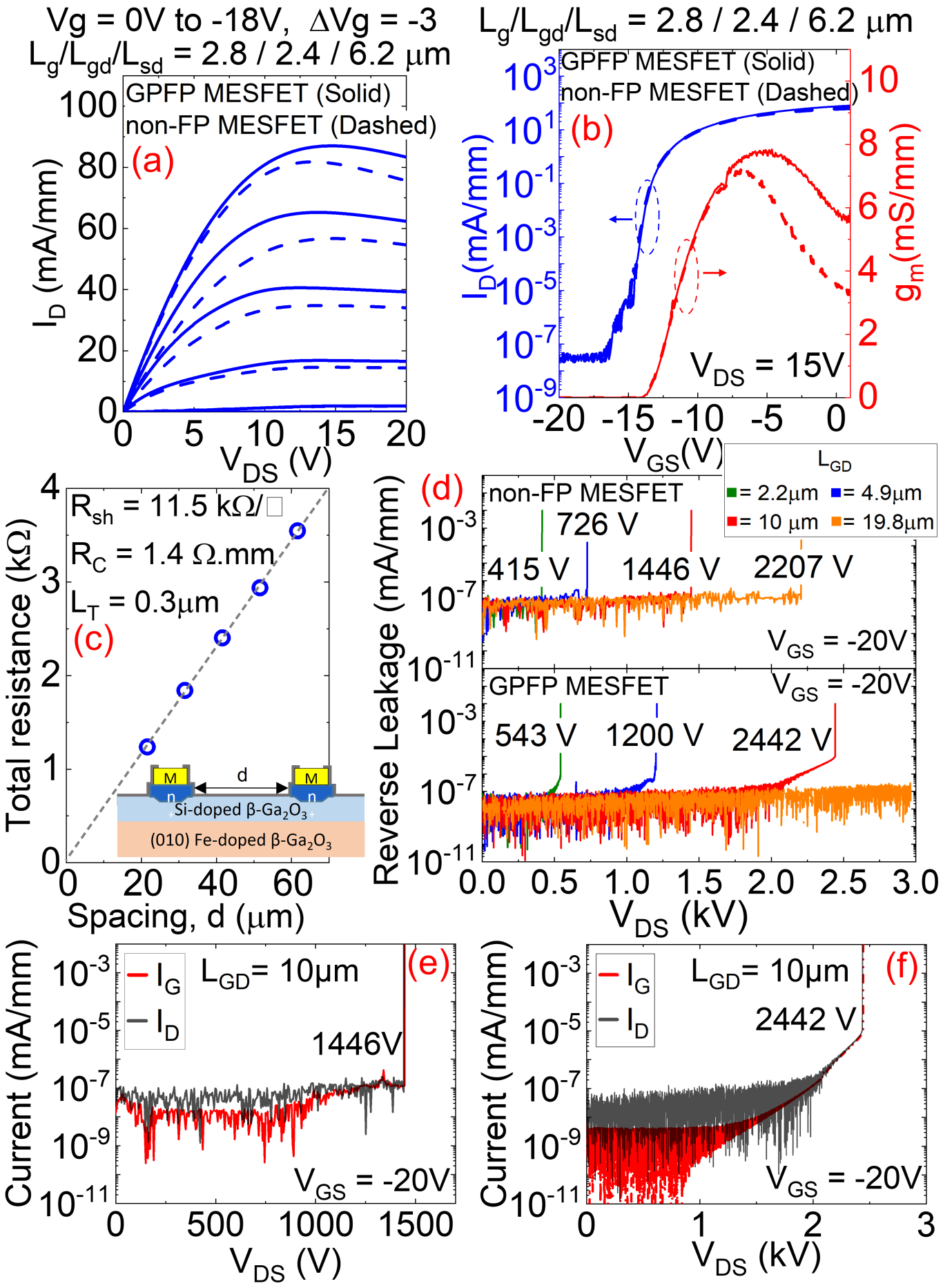

Fig. 2. (a) Output and (b) transfer curves for the GPFP (solid lines) and non-FP (dashed lines) β-$Ga_2O_3$ MESFETs. (c) Total resistance vs spacing for the TLM patterns (schematic shown in inset). (d) Breakdown characteristics of the non-FP (top) and GPFP (bottom) β-$Ga_2O_3$ MESFETs with various $L_{GD}$ values. Reverse leakage plots for the (e) non-FP MESFET with $L_{GD}$ of 10 µm and (f) GPFP MESFET with $L_{GD}$ of 10 µm.

It can be seen that for the GPFP MESFET, the $V_{BR}$ scales more linearly with $L_{GD}$ compared to the non-FP MESFETs (Fig.3(a)). Even though the E-field profile is highly non-uniform in the gate-drain region of a lateral device, an effective average lateral field ($E_{AVG} = V_{BR}/L_{GD}$) can be defined to be used as a metric to understand the scaling of breakdown voltage vs spacing and helps with lateral device design. Non-linear scaling of field peaking around the contact edges and transition from punch through to non-punch field profile occurs with increasing $L_{GD}$. This reduces $E_{BR,AVG}$ values and, therefore, results in non-linear $V_{BR}$ vs $L_{GD}$ relationship. However, the calculated $E_{AVG}$ values remained uniform ~2.5 MV/cm for GPFP MESFETs with $L_{GD}$ up to 10 µm. This combination of $E_{AVG}$ (~2.5 MV/cm) and $V_{BR}$ (~2.5 kV) demonstrated here for a MESFET with $L_{GD}$ = 10 µm is the highest to date for any depletion-mode $Ga_2O_3$ lateral device. GPFP MESFETs with $L_{GD}$ 20 µm and above did not breakdown and were measured repeatedly up to 3 kV (our measurement tool limit) without showing any degradation, demonstrating the robustness of these devices to high voltage stresses. It should be noted that all the devices in this work exhibit catastrophic breakdown.

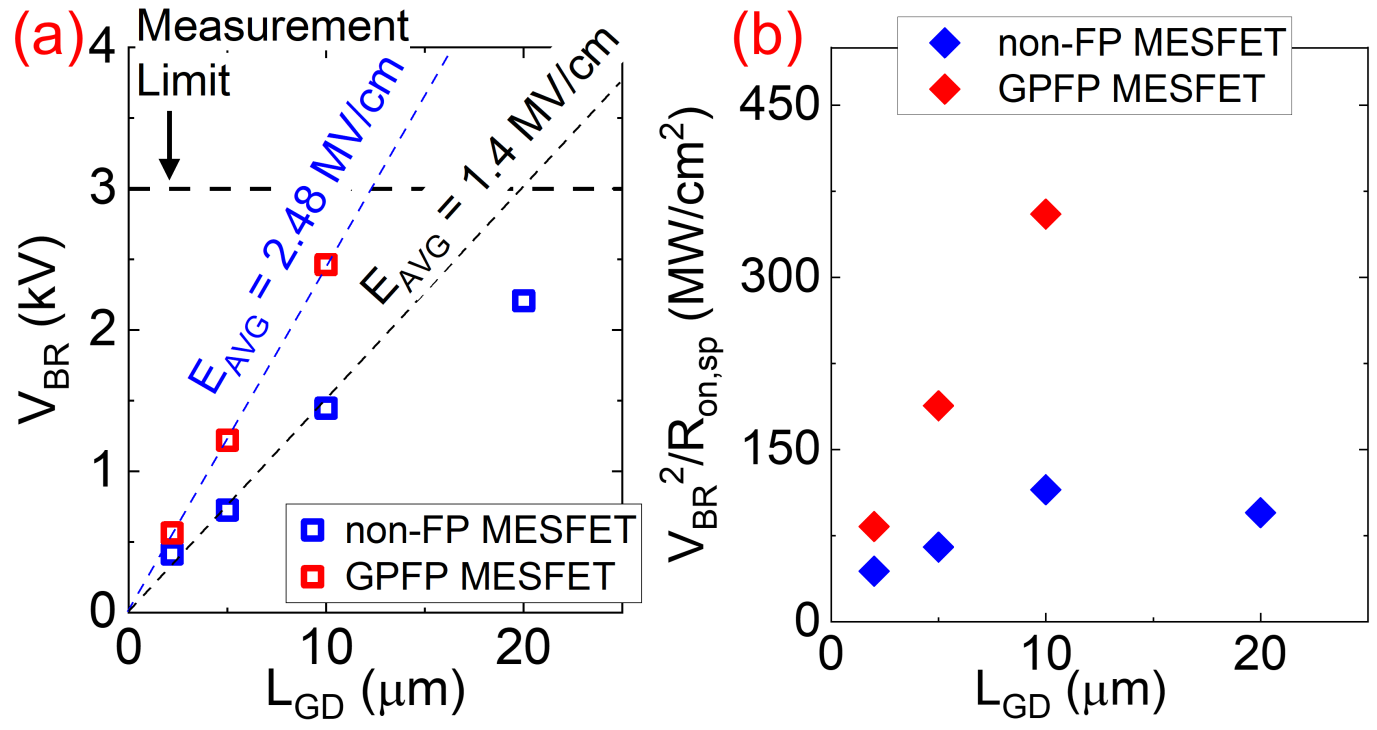

Fig. 3. (a) $V_{BR}$ as a function of $L_{GD}$ for the non-FP and GPFP MESFETs showing the increase in $E_{BR,AVG}$, and improved $V_{BR}$ vs $L_{GD}$ linearity. (b) LFOM of the MESFETs as a function of $L_{GD}$.

The lateral figures of merit ($V_{BR}^2/R_{on,sp}$) of the GPFP MESFETs were calculated, where $R_{on,sp}$ is $R_{ON}$ normalized to the device length ($L_{SD}+2L_T$). The LFOM values for the GPFP and non-FP MESFETs are plotted as a function of $L_{GD}$ (Fig.3(b)). The highest LFOM of 355 MW/cm$^2$ was calculated for GPFP MESFET ($V_{BR}$ = 2462V, $R_{on,sp}$ = 17.05 mΩ.cm$^2$) with an $L_{GD}$ of 10 µm which is more than 3× higher compared to the non-FP MESFET (LFOM = 115 MW/cm$^2$, $V_{BR}$ = 1466V, $R_{on,sp}$ = 18.6 mΩ.cm$^2$). Overall, the GPFP design is found to have 2-3 times higher LFOM compared to the non-FP design. This is because of the improved $R_{on,sp}$ (up to 14% lower) and $V_{BR}$ (up to 70% higher) values simultaneously with the GPFP design. Given the GPFP MESFET with $L_{GD}$ = 20 µm have $V_{BR}$ >3 kV, these devices can demonstrate LFOM >200 MW/cm$^2$ for >3kV class transistors.

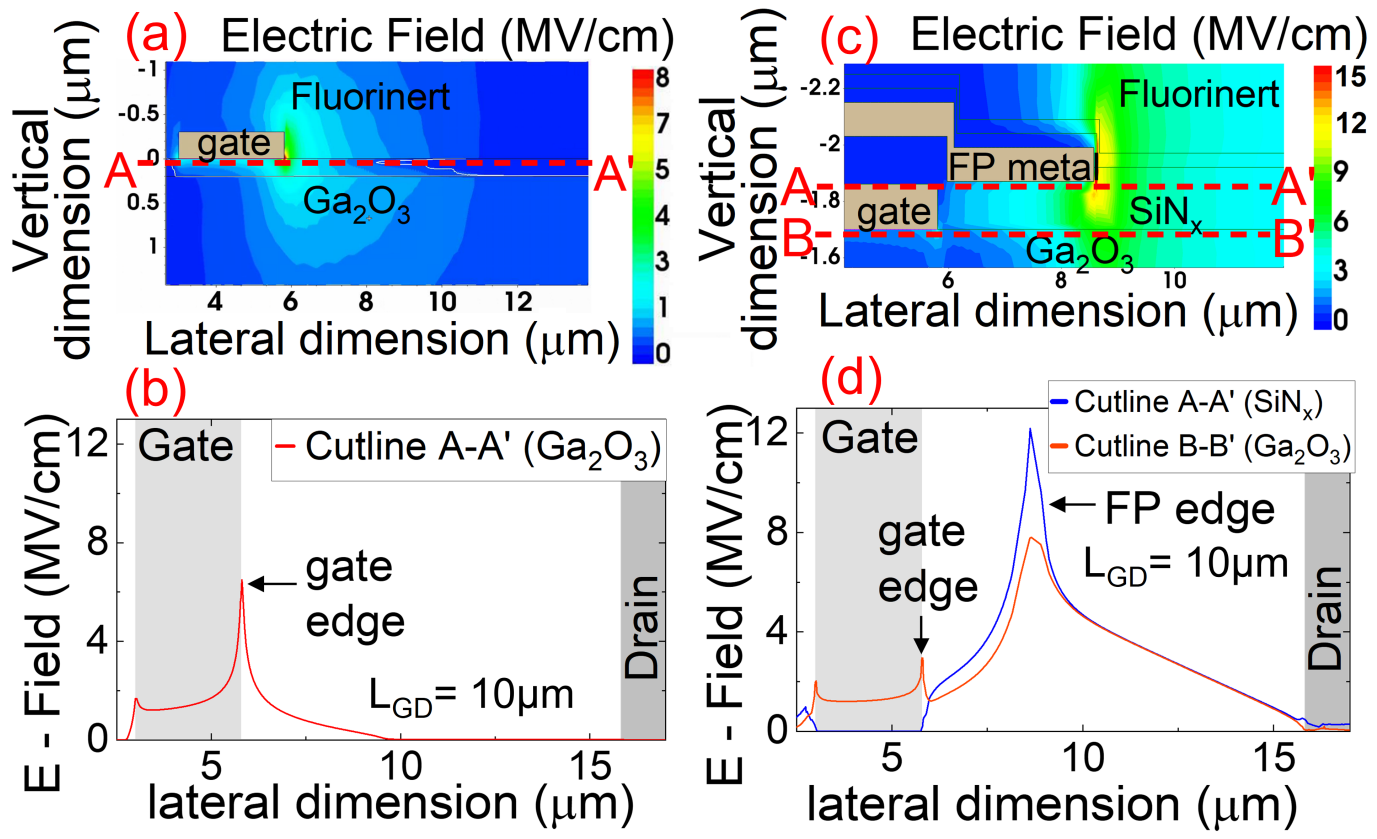

Fig. 4 (a) Simulated 2D electric field contour and (b) Lateral E-field profile in $Ga_2O_3$ under the gate along cutline A-A' and for a device with $L_{GD}$ = 10µm at the experimental breakdown condition of $V_{DG}$ = 1466V for the non-FP MESFET. (c) Simulated 2D electric field contour and (d) Lateral E-field profile along cutline A-A' & B-B' for a device with $L_{GD}$ = 10µm at the experimental breakdown condition of $V_{DG}$ = 2462V for the GPFP MESFET.

To understand the electric field profiles, the device structures were simulated using Sentaurus TCAD under the breakdown condition. For the GPFP design, the $E_{peak}$ was found to be at the FP edge as expected (Fig. 4(d)). The sharp metal edges used in the simulation act as points of singularity in the Poisson solver and may overestimate actual peak field values in real devices. From the simulated lateral field profile under the gate, the GPFP MESFET with an $L_{GD}$ of 10 µm was found to have a punchthrough field profile for the reported breakdown voltage.The electric field under the gate was found to be < 2 MV/cm, but the $E_{peak}$ values in the $SiN_x$ at the FP edge were found to be much higher (1.4 – 1.8× higher) than in $Ga_2O_3$ under the FP edge. Therefore, it is possible that the dielectric breakdown could also be the dominant breakdown mechanism for large $L_{GD}$ values rather than just the gate (tunneling) leakage in the GPFP design.

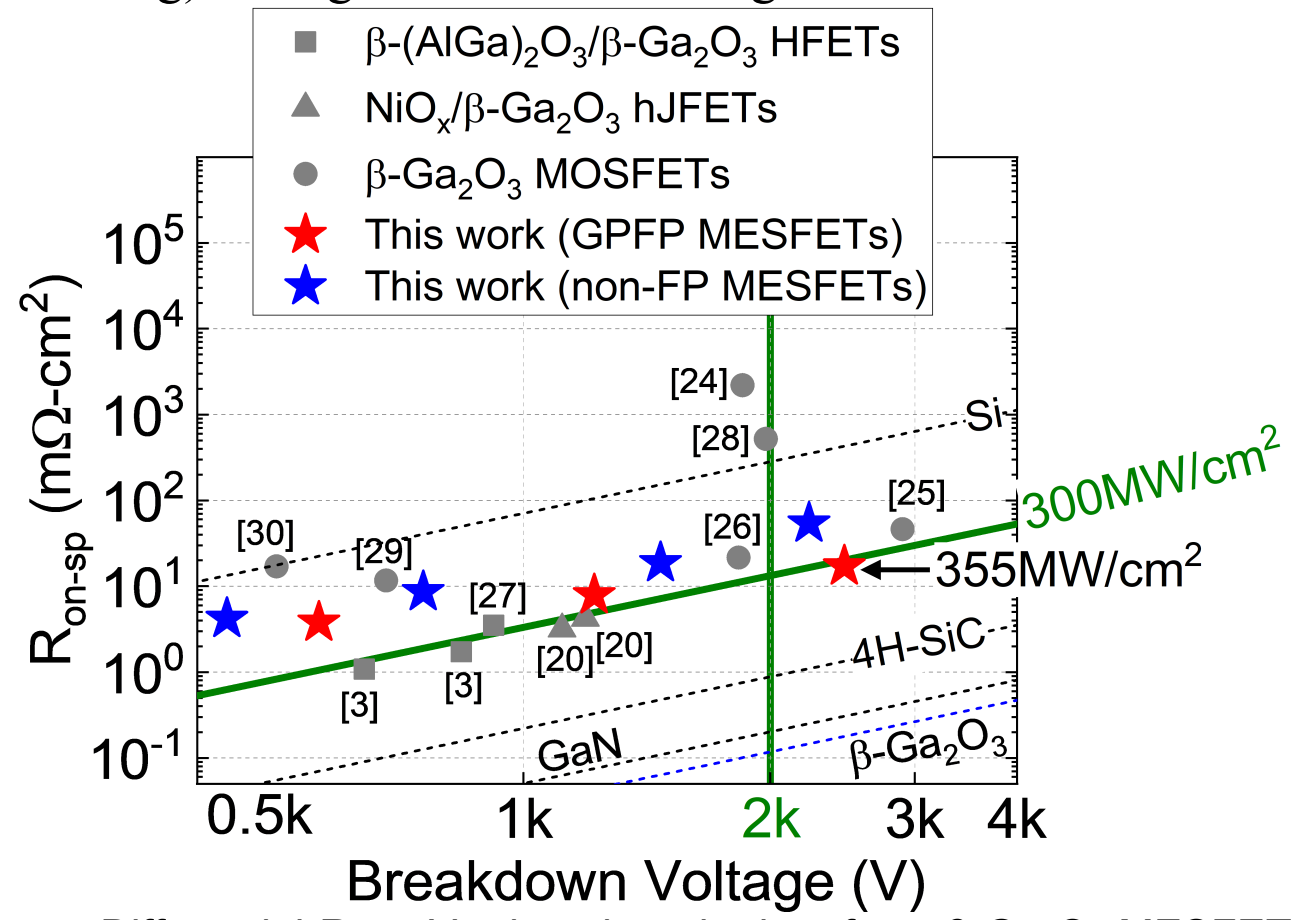

Fig. 5: Differential $R_{on,sp}$-$V_{BR}$ benchmark plot of our β-$Ga_2O_3$ MESFETs with the lateral β-$Ga_2O_3$ FETs in the literature [3], [20], [24]–[30].

Fig. 5 benchmarks our $V_{BR}$-$R_{on,sp}$ values with the existing literature reports showing that the reported GPFP β-$Ga_2O_3$ MESFET shows the highest LFOM for $V_{BR}$ > 2 kV, outperforming the likes of $Ga_2O_3$ MOSFETs and even advanced device designs like $(AlGa)_2O_3/Ga_2O_3$ HFETs and $NiO_x/Ga_2O_3$ p-n hetero-Junction FETs (hJFETs) [3], [20]. With the implementation of a gate dielectric, we expect the LFOM values could be further improved, provided gate dielectrics with low-leakage, cleaner interface and more importantly, dielectrics with high breakdown fields could be deposited along with high-quality β-$Ga_2O_3$ epitaxial layers.

## IV. Conclusion

We demonstrate gate-pad connected field plated MOVPE-grown kV-class β-$Ga_2O_3$ lateral MESFETs with high LFOM and $E_{BR,AVG}$ using PECVD deposited $SiN_x$ and $SiO_2$ field/passivation dielectrics. A record high LFOM of 355 MW/cm$^2$ with a $V_{BR}$ of ~2.5 kV and $E_{AVG}$ of ~2.5 MV/cm simultaneously is demonstrated in a GPFP β-$Ga_2O_3$ MESFET with $L_{GD}$ = 10 µm. This LFOM value is the highest for any β-$Ga_2O_3$ lateral device with $V_{BR}$ > 2kV. These devices show great potential of MOVPE-grown β-$Ga_2O_3$ FETs for future power device applications in the low to medium voltage range.

## REFERENCES

[1] M. Higashiwaki and G. H. Jessen, "Guest Editorial: The dawn of gallium oxide microelectronics," *Appl. Phys. Lett.*, vol. 112, no. 6, p. 060401, Feb. 2018, doi: 10.1063/1.5017845.

[2] S. J. Pearton, J. Yang, P. H. Cary, F. Ren, J. Kim, M. J. Tadjer, and M. A. Mastro, "A review of $Ga_2O_3$ materials, processing, and devices," *Applied Physics Reviews*, vol. 5, no. 1, p. 011301, Mar. 2018, doi: 10.1063/1.5006941.

[3] N. K. Kalarickal, Z. Xia, H.-L. Huang, W. Moore, Y. Liu, M. Brenner, J. Hwang, and S. Rajan, "β-$(Al_{0.18}Ga_{0.82})_2O_3$/$Ga_2O_3$ Double Heterojunction Transistor with Average Field of 5.5 MV/cm," *IEEE Electron Device Letters*, pp. 1–1, 2021, doi: 10.1109/LED.2021.3072052.

[4] S. Sharma, K. Zeng, S. Saha, and U. Singisetti, "Field-Plated Lateral $Ga_2O_3$ MOSFETs With Polymer Passivation and 8.03 kV Breakdown Voltage," *IEEE Electron Device Letters*, vol. 41, no. 6, pp. 836–839, Jun. 2020, doi: 10.1109/LED.2020.2991146.

[5] S. Roy, A. Bhattacharyya, P. Ranga, H. Splawn, J. Leach, and S. Krishnamoorthy, "High-k Oxide Field-Plated Vertical (001) β- $Ga_2O_3$ Schottky Barrier Diode with Baliga's Figure of Merit Over 1 GW/cm2," *IEEE Electron Device Letters*, pp. 1–1, 2021, doi: 10.1109/LED.2021.3089945.

[6] A. J. Green, K. D. Chabak, E. R. Heller, R. C. Fitch, M. Baldini, A. Fiedler, K. Irmscher, G. Wagner, Z. Galazka, S. E. Tetlak, A. Crespo, K. Leedy, and G. H. Jessen, "3.8-MV/cm Breakdown Strength of MOVPE-Grown Sn-Doped β-$Ga_2O_3$ MOSFETs," *IEEE Electron Device Letters*, vol. 37, no. 7, pp. 902–905, Jul. 2016, doi: 10.1109/LED.2016.2568139.

[7] K. Sasaki, M. Higashiwaki, A. Kuramata, T. Masui, and S. Yamakoshi, "MBE grown $Ga_2O_3$ and its power device applications," *Journal of Crystal Growth*, vol. 378, pp. 591–595, 2013.

[8] Z. Feng, A. F. M. Anhar Uddin Bhuiyan, M. R. Karim, and H. Zhao, "MOCVD homoepitaxy of Si-doped (010) β-$Ga_2O_3$ thin films with superior transport properties," *Appl. Phys. Lett.*, vol. 114, no. 25, p. 250601, Jun. 2019, doi: 10.1063/1.5109678.

[9] S. Bin Anooz, R. Grüneberg, C. Wouters, R. Schewski, M. Albrecht, A. Fiedler, K. Irmscher, Z. Galazka, W. Miller, G. Wagner, J. Schwarzkopf, and A. Popp, "Step flow growth of β-$Ga_2O_3$ thin films on vicinal (100) β-$Ga_2O_3$ substrates grown by MOVPE," *Appl. Phys. Lett.*, vol. 116, no. 18, p. 182106, May 2020, doi: 10.1063/5.0005403.

[10] J. H. Leach, K. Udwary, J. Rumsey, G. Dodson, H. Splawn, and K. R. Evans, "Halide vapor phase epitaxial growth of β-$Ga_2O_3$ and α-$Ga_2O_3$ films," *APL Materials*, vol. 7, no. 2, p. 022504, Feb. 2019, doi: 10.1063/1.5055680.

[11] H. Murakami, K. Nomura, K. Goto, K. Sasaki, K. Kawara, Q. T. Thieu, R. Togashi, Y. Kumagai, M. Higashiwaki, A. Kuramata, S. Yamakoshi, B. Monemar, and A. Koukitu, "Homoepitaxial growth of β-$Ga_2O_3$ layers by halide vapor phase epitaxy," *Appl. Phys. Express*, vol. 8, no. 1, p. 015503, Dec. 2014, doi: 10.7567/APEX.8.015503.

[12] P. Ranga, A. Rishinaramangalam, J. Varley, A. Bhattacharyya, D. Feezell, and S. Krishnamoorthy, "Si-doped β-$(Al_{0.26}Ga_{0.74})_2O_3$ thin films and heterostructures grown by metalorganic vapor-phase epitaxy," *Applied Physics Express*, vol. 12, no. 11, p. 111004, Oct. 2019, doi: 10.7567/1882-0786/ab47b8.

[13] P. Ranga, A. Bhattacharyya, A. Rishinaramangalam, Y. K. Ooi, M. A. Scarpulla, D. Feezell, and S. Krishnamoorthy, "Delta-doped β-$Ga_2O_3$ thin films and β-$(Al_{0.26}Ga_{0.74})_2O_3$/β-$Ga_2O_3$ heterostructures grown by metalorganic vapor-phase epitaxy," *Appl. Phys. Express*, vol. 13, no. 4, p. 045501, Mar. 2020, doi: 10.35848/1882-0786/ab7712.

[14] P. Ranga, A. Bhattacharyya, A. Chmielewski, S. Roy, N. Alem, and S. Krishnamoorthy, "Delta-doped β-$Ga_2O_3$ films with narrow FWHM grown by metalorganic vapor-phase epitaxy," *Applied Physics Letters*, vol. 117, no. 17, p. 172105, 2020, doi: 10.1063/5.0027827.

[15] P. Ranga, A. Bhattacharyya, A. Chmielewski, S. Roy, R. Sun, M. A. Scarpulla, N. Alem, and S. Krishnamoorthy, "Growth and characterization of metalorganic vapor-phase epitaxy-grown β-$(Al_xGa_{1-x})_2O_3$/β-$Ga_2O_3$ heterostructure channels," *Appl. Phys. Express*, vol. 14, no. 2, p. 025501, Jan. 2021, doi: 10.35848/1882-0786/abd675.

[16] P. Ranga, A. Bhattacharyya, L. Whittaker-Brooks, M. A. Scarpulla, and S. Krishnamoorthy, "N-type doping of low-pressure chemical vapor deposition grown β-$Ga_2O_3$ thin films using solid-source germanium," *Journal of Vacuum Science & Technology A*, vol. 39, no. 3, p. 030404, May 2021, doi: 10.1116/6.0001004.

[17] A. Bhattacharyya, P. Ranga, S. Roy, J. Ogle, L. Whittaker-Brooks, and S. Krishnamoorthy, "Low temperature homoepitaxy of (010) β-$Ga_2O_3$ by metalorganic vapor phase epitaxy: Expanding the growth window," *Appl. Phys. Lett.*, vol. 117, no. 14, p. 142102, Oct. 2020, doi: 10.1063/5.0023778.

[18] Y. Zhang, F. Alema, A. Mauze, O. S. Koksaldi, R. Miller, A. Osinsky, and J. S. Speck, "MOCVD grown epitaxial β-$Ga_2O_3$ thin film with an electron mobility of 176 cm2/V s at room temperature," *APL Materials*, vol. 7, no. 2, p. 022506, Dec. 2018, doi: 10.1063/1.5058059.

[19] G. Seryogin, F. Alema, N. Valente, H. Fu, E. Steinbrunner, A. T. Neal, S. Mou, A. Fine, and A. Osinsky, "MOCVD growth of high purity $Ga_2O_3$ epitaxial films using trimethylgallium precursor," *Appl. Phys. Lett.*, vol. 117, no. 26, p. 262101, Dec. 2020, doi: 10.1063/5.0031484.

[20] C. Wang, H. Gong, W. Lei, Y. Cai, Z. Hu, S. Xu, Z. Liu, Q. Feng, H. Zhou, J. Ye, J. Zhang, R. Zhang, and Y. Hao, "Demonstration of the p-$NiO_x$/n-$Ga_2O_3$ Heterojunction Gate FETs and Diodes With $BV^2/R_{on,sp}$ Figures of Merit of 0.39 GW/$cm^2$ and 1.38 GW/$cm^2$," *IEEE Electron Device Letters*, vol. 42, no. 4, pp. 485–488, Apr. 2021, doi: 10.1109/LED.2021.3062851.

[21] A. Bhattacharyya, S. Roy, P. Ranga, D. Shoemaker, Y. Song, J. S. Lundh, S. Choi, and S. Krishnamoorthy, "130 mA mm-1 β-$Ga_2O_3$ metal semiconductor field effect transistor with low-temperature metalorganic vapor phase epitaxy-regrown ohmic contacts," *Appl. Phys. Express*, vol. 14, no. 7, p. 076502, Jun. 2021, doi: 10.35848/1882-0786/ac07ef.

[22] C. Joishi, Z. Xia, J. S. Jamison, S. H. Sohel, R. C. Myers, S. Lodha, and S. Rajan, "Deep-Recessed β-$Ga_2O_3$ Delta-Doped Field-Effect Transistors With In Situ Epitaxial Passivation," *IEEE Transactions on Electron Devices*, vol. 67, no. 11, pp. 4813–4819, Nov. 2020, doi: 10.1109/TED.2020.3023679.

[23] C. Joishi, Y. Zhang, Z. Xia, W. Sun, A. R. Arehart, S. Ringel, S. Lodha, and S. Rajan, "Breakdown Characteristics of β-$(Al_{0.22}Ga_{0.78})_2O_3$/$Ga_2O_3$ Field-Plated Modulation-Doped Field-Effect Transistors," *IEEE Electron Device Letters*, vol. 40, no. 8, pp. 1241–1244, Aug. 2019, doi: 10.1109/LED.2019.2921116.

[24] K. Zeng, A. Vaidya, and U. Singisetti, "1.85 kV Breakdown Voltage in Lateral Field-Plated $Ga_2O_3$ MOSFETs," *IEEE Electron Device Letters*, vol. 39, no. 9, pp. 1385–1388, Sep. 2018, doi: 10.1109/LED.2018.2859049.

[25] Y. Lv, H. Liu, X. Zhou, Y. Wang, X. Song, Y. Cai, Q. Yan, C. Wang, S. Liang, J. Zhang, Z. Feng, H. Zhou, S. Cai, and Y. Hao, "Lateral β-$Ga_2O_3$ MOSFETs With High Power Figure of Merit of 277 MW/$cm^2$," *IEEE Electron Device Letters*, vol. 41, no. 4, pp. 537–540, Apr. 2020, doi: 10.1109/LED.2020.2974515.

[26] K. Tetzner, E. Bahat Treidel, O. Hilt, A. Popp, S. Bin Anooz, G. Wagner, A. Thies, K. Ickert, H. Gargouri, and J. Würfl, "Lateral 1.8 kV β-$Ga_2O_3$ MOSFET With 155 MW/$cm^2$ Power Figure of Merit," *IEEE Electron Device Letters*, vol. 40, no. 9, pp. 1503–1506, Sep. 2019, doi: 10.1109/LED.2019.2930189.

[27] N. K. Kalarickal, Z. Feng, A. F. M. Anhar Uddin Bhuiyan, Z. Xia, W. Moore, J. F. McGlone, A. R. Arehart, S. A. Ringel, H. Zhao, and S. Rajan, "Electrostatic Engineering Using Extreme Permittivity Materials for Ultra-Wide Bandgap Semiconductor Transistors," *IEEE Transactions on Electron Devices*, vol. 68, no. 1, pp. 29–35, Jan. 2021, doi: 10.1109/TED.2020.3037271.

[28] K. Zeng, A. Vaidya, and U. Singisetti, "A field-plated $Ga_2O_3$ MOSFET with near 2-kV breakdown voltage and 520 mΩ.$cm^2$," *Appl. Phys. Express*, vol. 12, no. 8, p. 081003, Jul. 2019, doi: 10.7567/1882-0786/ab2e86.

[29] Y. Lv, X. Zhou, S. Long, X. Song, Y. Wang, S. Liang, Z. He, T. Han, X. Tan, Z. Feng, H. Dong, X. Zhou, Y. Yu, S. Cai, and M. Liu, "Source-Field-Plated β-$Ga_2O_3$ MOSFET With Record Power Figure of Merit of 50.4 MW/$cm^2$," *IEEE Electron Device Letters*, vol. 40, no. 1, pp. 83–86, Jan. 2019, doi: 10.1109/LED.2018.2881274.

[30] K. D. Chabak, J. P. McCandless, N. A. Moser, A. J. Green, K. Mahalingam, A. Crespo, N. Hendricks, B. M. Howe, S. E. Tetlak, K. Leedy, R. C. Fitch, D. Wakimoto, K. Sasaki, A. Kuramata, and G. H. Jessen, "Recessed-Gate Enhancement-Mode β-$Ga_2O_3$ MOSFETs," *IEEE Electron Device Letters*, vol. 39, no. 1, pp. 67–70, Jan. 2018, doi: 10.1109/LED.2017.2779867.